\documentstyle[preprint,tighten,aps,epsfig,psfrag]{revtex}
\begin{document}

\newcommand{\TeV}{\,{\rm TeV}}
\newcommand{\GeV}{\,{\rm GeV}}
\newcommand{\MeV}{\,{\rm MeV}}
\newcommand{\keV}{\,{\rm keV}}
\newcommand{\eV}{\,{\rm eV}}
\def\beqar{\begin{eqnarray}}
\def\eeqar{\end{eqnarray}}
\def\beq{\begin{equation}}
\def\eeq{\end{equation}}
\def\haf{\frac{1}{2}}
\def\plb#1#2#3#4{#1, Phys. Lett. {\bf #2B} (#4) #3}
\def\plbb#1#2#3#4{#1 Phys. Lett. {\bf #2B} (#4) #3}
\def\npb#1#2#3#4{#1, Nucl. Phys. {\bf B#2} (#4) #3}
\def\prd#1#2#3#4{#1, Phys. Rev. {\bf D#2} (#4) #3}
\def\prl#1#2#3#4{#1, Phys. Rev. Lett. {\bf #2} (#4) #3}
\def\mpl#1#2#3#4{#1, Mod. Phys. Lett. {\bf A#2} (#4) #3}
\def\rep#1#2#3#4{#1, Phys. Rep. {\bf #2} (#4) #3}
\def\lpp{\lambda''}
\def\ccg{\cal G}
\def\slash#1{#1\!\!\!\!\!/}
\def\rpv{\slash{R_p}}
\def\fui{f^{(u)}_i}
\def\fuj{f^{(u)}_j}
\def\fdi{f^{(d)}_i}
\def\fdj{f^{(d)}_j}
\def\mui{m^{(u)}_i}
\def\muj{m^{(u)}_j}
\def\mdi{m^{(d)}_i}
\def\mdj{m^{(d)}_j}
 
\draft
\preprint{
\begin{tabular}{r}
KAIST-TH-98/02
\\
hep-ph/yymmddd
\end{tabular}
}
\title{
Constraints  on $R$ parity and $B$ violating couplings
in gauge-mediated supersymmetry breaking models
}
\author{
Kiwoon Choi
\thanks{E-mail: kchoi@chep6.kaist.ac.kr},
Kyuwan Hwang
\thanks{E-mail: kwhwang@supy.kaist.ac.kr},
and
Jae Sik Lee
\thanks{E-mail: jslee@chep6.kaist.ac.kr}
}
\address{
Department of Physics, Korea Advanced Institute of Science and
Technology \\
Taejon 305-701, Korea \\
}

\maketitle

\begin{abstract}
We consider 
the proton decay involving a light gravitino
or axino in gauge-mediated supersymmetry 
breaking models to derive
constraints on the $R$ parity and baryon number violating
Yukawa couplings.
Bounds on all nine coupling constants  are obtained by considering
the decay amplitudes at one-loop order.

\end{abstract}



In supersymmetric models,
there can be renormalizable 
gauge-invariant terms in the superpotential  which violate
the baryon number $B$ or the lepton number $L$.
To avoid such terms, one usually introduces an additional discrete
symmetry, 
the so called $R$ parity ($R_p=(-1)^{3B+L+2S}$).
Although  $R_p$ conservation leads to a consistent theory,
there is no compelling  theoretical reason to assume
this symmetry.
It is therefore an interesting possibility
to have  an explicit 
$R_p$ violation which may lead to interesting phenomenological
consequences
\cite{ago}.
In the minimal supersymmetric standard model,
the most general $R_p$-violating superpotential is given by
\beq
\frac{1}{2}\lambda_{ijk}L_iL_jE_k^c+\lambda_{ijk}'L_iQ_jD_k^c+
\frac{1}{2}\lambda_{ijk}''U_i^cD_j^cD_k^c,
\label{superW}
\eeq
where
$L_i$ and $Q_i$ are the $SU(2)$-doublet lepton and quark superfields and
$E_i^c,U_i^c,D_i^c$ are the singlet superfields, respectively.
Here $i,j,k$ are generation indices and we assume that possible bilinear terms
$\mu_i L_i H_2$ are rotated away.
Obviously the first and second terms in (\ref{superW}) violate $L$, while
the third violates $B$.
Since 
$\lambda_{ijk}=-\lambda_{jik}$ and
$\lambda_{ijk}''=-\lambda_{ikj}''$, $R_p$ violations are described
by 45 complex Yukawa couplings
(9 in 
$\lambda$, 27 in $\lambda'$ and  9 in $\lambda''$).

It is well known that the consideration of proton decay
provides a very stringent constraint on the product of
$\lambda'$ and $\lambda''$:
\beq
\left| \lambda'_{112}\lambda''_{112}\right| ,
\left| \lambda'_{123}\lambda''_{113}\right| \leq 10^{-24},
\eeq
where the squark masses are assumed to be around 1 TeV \cite{hinkae}.
This bound  has been obtained from a
squark-mediated proton decay at {\it tree} level which does not 
involve heavy generation particles and thus applies
for the particular combination of generation indices
as is shown above.
One may then expect that other products of $\lambda'$
and $\lambda''$ 
are allowed to be large.
However it has been noted \cite{smivi} that 
for any pair of $\lambda'$ and $\lambda''$, 
there is always at least
one diagram relevant for the proton decay at one-loop level. This means
that all products of $\lambda'$ and $\lambda''$ can be constrained
by proton decay and a more detailed analysis leads to \cite{smivi}
\beq
\left| \lambda'\cdot\lambda''\right| \leq 10^{-9},
\eeq
for {\it any} pair of $\lambda'$ and $\lambda''$.

As was noted in \cite{chang,ccl},  
if there is a light fermion (lighter than the proton)
which does {\it not} carry any lepton number,
proton decay can be induced
by  a $B$ violating but $L$ conserving
interaction alone, for instance
by the $\lambda''$ couplings alone.
Perhaps the most interesting class of models
predicting such a light fermion are
supersymmetric models in which supersymmetry (SUSY)
breaking is mediated by  gauge interactions \cite{GM}.
In such models,
the squark and/or gaugino masses, i.e. the soft masses in the supersymmetric
standard model (SSM) sector,
are given by
$m_{\rm soft}\simeq ({\alpha\over \pi})^n \Lambda_S$
where $\Lambda_S$ corresponds to the scale of spontaneous
SUSY breaking and 
the model-dependent integer $n$ counts the number of loops
involved in transmitting SUSY breaking to the supersymmetric
standard model sector. 
On the other hands,
the gravitino mass is suppressed by the Planck scale
$M_P\simeq 2\times 10^{18}$ GeV,
$m_{3/2}=\Lambda^2_S/M_P$.
Assuming  that $m_{\rm soft}$ is at the weak scale and taking
$n=1\sim 3$ for instance,
we have $m_{3/2}\simeq 10^{-1}\eV \sim 10 \MeV$
which is  far below the proton mass.

Another interesting candidate for a light fermion without carrying any lepton
number 
is the axino in supersymmetric models with  a spontaneously broken 
global  $U(1)_{PQ}$ 
symmetry \cite{pq}.
If SUSY breaking is  mediated by gauge interactions,
the axino mass is given by
$m_{\tilde{a}}\simeq (\alpha/\pi)^m\Lambda_S^2/F_a$ where
$m$ is again a model-dependent  (but typically {\it not} less than $n$)
integer and  $F_a$ denotes the scale
of spontaneous $U(1)_{PQ}$ breaking \cite{axino}.
Obviously 
in this case the axino can be  lighter than the proton
for  a  phenomenologically allowed $F_a\geq 10^{10}$ GeV.
In other type of models in which SUSY breaking is transmitted
by supergravity interactions,  the gravitino mass is fixed to be
of the weak scale order, however there is still a room for an axino
lighter than the proton \cite{chun}.
As was pointed out in Ref. \cite{chun},
some supergravity-mediated models lead to
$m_{\tilde{a}}\simeq m_{3/2}(m_{3/2}/M_P)^{1/2}\simeq 1$ keV for which
the axino would be a good warm dark matter candidate \cite{raja}.

\begin{figure}[h]
\hspace{1pt}
\vspace{75pt}
\begin{center}
\epsfig{file=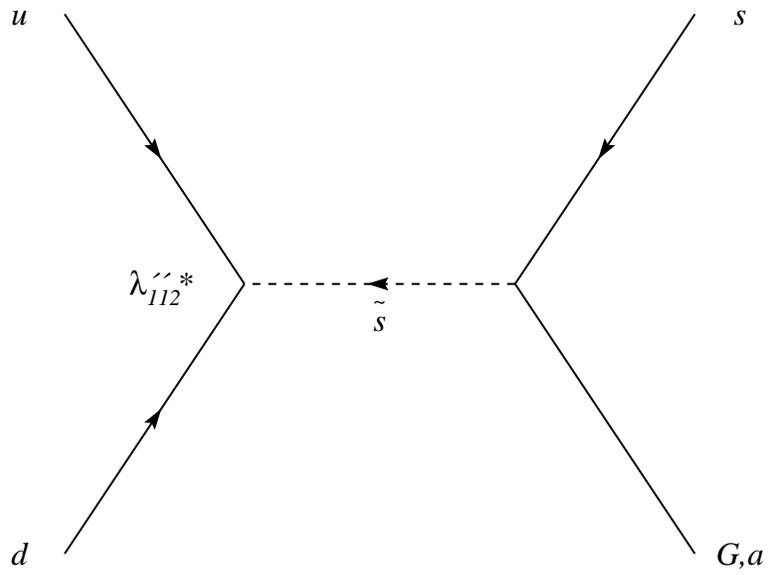}
\end{center}
\caption{
Tree diagram for the proton decay into light  gravitino
or axino. 
\label{fig1}
}
\end{figure}

In Ref. \cite{ccl},
proton decay involving  a light gravitino or axino  
has been analysed at tree approximation to obtain
a constraint on the $R_p$ and $B$ violating Yukawa coupling
$\lambda''_{ijk}$.
Applying the naive dimensional analysis rule \cite{nda} for the hadronic
matrix element of the effective 4-fermion operator 
induced by the tree diagram of Fig.\ref{fig1}, 
the following stringent bounds on 
$\lambda_{112}''$ (in the quark mass eigenstate basis) were obtained:
\beqar
\lambda^{\prime\prime}_{112}&\leq& 5\times 10^{-16}
\left(\frac{\tilde{m}}{300 \, {\rm GeV}}\right)^2
\left(\frac{m_{3/2}}{1 \, {\rm eV}}\right) \, , \nonumber \\
\lambda^{\prime\prime}_{112}&\leq& 7\times 10^{-16} 
\left(\frac{\tilde{m}}{300 \, {\rm GeV}}\right)^2
\left(\frac{F_a}{10^{10} \, {\rm GeV}}\right) \ 
\left(\frac{1}{c_q}\right) \ , 
\label{treeconstraint}
\eeqar
where $\tilde{m}$ denotes the squark mass which is presumed to
be universal. Here the dimensionless  coefficient $c_q$ describes
the axino coupling to
the light quarks $q=(u,d,s)$ and 
is of order one for Dine-Fischler-Srednicki-Zhitnitskii
type axino, while it is of order $10^{-2}\sim 10^{-3}$
for hadronic-type axino.
In this paper, we wish to extend the analysis of \cite{ccl}
by including one-loop effects and  derive the constraints 
on the other components of $\lambda''$.
The present upper bounds on  $\lpp$ are ${\cal O}(1)$ for
sfermion mass $\tilde{m} \sim 100$ GeV except those 
on $\lambda''_{112}$  and $\lambda''_{113}$
\cite{bha}.
As we will see,  the derived  upper bounds on $\lpp_{ijk}$ in gauge-mediated
SUSY breaking models 
are much stronger than
the currently existing bounds for a wide range of $m_{3/2}$ and $F_a$.

At low energy scales $\sim 1$ GeV where all massive particles
are integrated out,
the proton decay
$p\rightarrow \psi +$ light meson ($\psi=$ gravitino or axino)
can be described by an effective 4-fermion operator,
${\cal O}_{\rm eff}=uds\psi$ or $udd\psi$,  in the quark mass eigenstate basis. 
(See \cite{ccl} for the  detailed kinematic structure  
of these 4-fermion operators, which is not essential for
our discussion in this paper.)
At tree approximation, only $\lambda''_{112}$ can produce
such an effective operator (see  Fig.\ref{fig1}), thereby  is constrained
as (\ref{treeconstraint}).
In order for the other $\lambda''_{ijk}$ to produce
the flavor structure 
$uds$ or $udd$, it must be supplemented
by flavor changing interactions in the model, which is 
possible at one-loop order.  
For instance,  $\lambda''_{123}$ and
$\lambda''_{113}$
can induce ${\cal O}_{\rm eff}$ once they
are combined with the flavor change $b\rightarrow d$,
while $\lambda''_{212}$
and $\lambda''_{312}$ can do it with 
the flavor changes $c\rightarrow u$ and $t\rightarrow u$, respectively.
The other four couplings need  double flavor changes in order
to lead to a proton decay, e.g.
$(c,b)\rightarrow (u, d)$ or $(u,s)$
for $\lambda''_{213}$
and  $\lambda''_{223}$, $(t,b)\rightarrow (u, d)$ or $(u,s)$
for $\lambda''_{313}$ 
and $\lambda''_{323}$.

To proceed, let us collect the couplings which are relevant for the
proton decay into  light gravitino or axino at one-loop order.  
First of all,  one needs the following gravitino ($G$) \cite{fayet}
or axino ($\tilde{a}$) \cite{kim}
couplings:
\beqar
{\cal L}_{G}&=&
\frac{i}{4\sqrt{6}m_{3/2}M_P}
\left[
\bar{\lambda}^{\alpha}\gamma^{\rho}\sigma^{\mu\nu}
\partial_{\rho}GF_{\mu\nu}^{\alpha}
+
2\sqrt{2} \bar{\psi}_{I}(1-\gamma_5)
\gamma^{\mu}\gamma^{\nu}
\partial_{\mu}GD_{\nu}\phi^*_I+ {\rm H.c.} \right],\nonumber \\
{\cal L}_{\tilde{a}}&=&
-\frac{c_{_I}}{2F_a} \left[i
\partial_{\mu}\bar{\psi}_I\gamma^{\mu}(1+\gamma_5)\tilde{a}\phi^*_I
+ {\rm H.c.} \right]  
+\frac{c_{\alpha}}{32\sqrt{2}\pi^2 F_a}
\left[\bar{\lambda}^{\alpha}
\gamma^{\mu}\gamma^{\nu}(1-\gamma_5)\tilde{a}F^{\alpha}_{\mu\nu}+
{\rm H.c.}\right],
\label{gacoupling}
\eeqar
where  $(\phi_I,\psi_I)$ and $(\lambda^{\alpha}, F^{\alpha}_{\mu\nu})$ stand for
the chiral matter and gauge multiplets.
Here the axino couplings $c_{\alpha}$ to gauge multiplets
are generically
of order one, while the couplings $c_{_I}$  to matter multiplets  are
of order one
only for matter multiplets carrying a nonzero
$U(1)_{PQ}$ charge \cite{kim}.
For matter multiplets with  vanishing
$U(1)_{PQ}$ charge,
$c_I$ are 
loop--suppressed and thus
of order $10^{-2}\sim 10^{-3}$.

The second relevant couplings are those associated with
the  $\lpp$-term  in the superpotential (\ref{superW}),
for instance
the following
$B$-violating quark-quark-squark Yukawa couplings:
\beq
  -\lambda''_{ijk} \left[
  \bar{u}^i P_L (d^j)^C \tilde{d}^{k*}_R +
\bar{d}^j P_L (d^k)^C \tilde{u}^{i*}_R
  \right] + {\rm H.c.},
\label{rpvL}
\eeq
where $P_{L,R} = (1\mp\gamma_5)/2$
and the upper indices of quarks and squarks  designate the generation
number.

As was noted above, except for the case of $\lambda''_{112}$,
one needs
flavor changing
interactions  to accomplish the proton decay induced by 
$\lambda''_{ijk}$.
In gauge-mediated SUSY breaking models,
flavor changing neutral current
interactions in $R_p$-conserving sector are highly
suppressed.
Then the necessary flavor change takes place through
the exchange of  $W^{\pm}$, charged Higgs or charginos.
The flavor changing interactions which we  will use in the 
subsequent discussions include the $W$-boson coupling:
\beq
-V_{ij} \frac{g}{\sqrt{2}} W_{\mu}^+ \bar{u}^i \gamma^{\mu} P_L d^j + {\rm H.c.},
\eeq
and the charged Higgs boson coupling:
\beq
V_{ij} H^+ \left[\fdj\tan\beta \, \bar{u}^i P_R d^j
+\fui\cot\beta \, \bar{u}^i P_L d^j \right]
+ {\rm H.c.},
\eeq
where $g$ is the $SU(2)$ gauge coupling, 
$V_{ij}$ is the CKM matrix element,
$\tan\beta$ is the ratio of Higgs vacuum
expectation values,
and $f^{(u,d)}_i$ denote the quark Yukawa couplings, i.e.
\beq
\fui=\frac{g\mui}{\sqrt{2}m_W\sin\beta},
~~~~
\fdi=\frac{g\mdi}{\sqrt{2}m_W\cos\beta}.
\label{yuk}
\eeq
(Here $m_W$, $\mui$ and $\mdi$  denote the masses of 
$W$-boson, up- and down-type quarks respectively.)
In addition to these, we will use the following 
chargino-quark-squark interactions also:
\beqar
 && V_{ij}^* \left\{
  \overline{\tilde{\chi}}_1^+ \left(
  -g\cos\phi_L P_L + \fui\sin\phi_R P_R \right) u^i\tilde{d}^{j*}_L +
  \fdj\sin\phi_L \overline{\tilde{\chi}}_1^+ P_L  u^i\tilde{d}^{j*}_R
        \right. \nonumber \\
  && ~~  \left. +\overline{\tilde{\chi}}_2^+ \left(
  g\sin\phi_L P_L + \fui\epsilon_R\cos\phi_R P_R \right) u^i\tilde{d}^{j*}_L
  + \fdj\cos\phi_L \overline{\tilde{\chi}}_2^+ P_L  u^i\tilde{d}^{j*}_R
    \right\} \nonumber \\
  && +V_{ij} \left\{
  \overline{\tilde{\chi}}_1^- \left(
  -g\cos\phi_R P_L + \fdj\sin\phi_L P_R \right) d^j\tilde{u}^{i*}_L +
  \fui\sin\phi_R \overline{\tilde{\chi}}_1^- P_L  d^j\tilde{u}^{i*}_R
        \right. \nonumber \\
  && ~~ \left. +\overline{\tilde{\chi}}_2^- \left(
  g\sin\phi_R P_L + \fdj\cos\phi_L P_R \right) d^j\tilde{u}^{i*}_L +
  \fui\epsilon_R\cos\phi_R \overline{\tilde{\chi}}_2^- P_L
d^j\tilde{u}^{i*}_R
    \right\} + {\rm H.c.},
\eeqar
where 
$\epsilon_R \equiv {\rm sign}(\mu M_2-m_W^2\sin2\beta)$
for the gaugino mass $M_2$ and the Higgsino mass parameter $\mu$. 
The chargino mixing angles $\phi_{L,R}$ are given by
\beqar
  \tan 2\phi_L = \frac{ 2\sqrt{2}m_W(M_2\cos\beta + \mu\sin\beta)}
       {M_2^2-\mu^2-2m_W^2\cos2\beta}, \nonumber \\
  \tan 2\phi_R = \frac{ 2\sqrt{2}m_W(M_2\sin\beta + \mu\cos\beta)}
       {M_2^2-\mu^2+2m_W^2\cos2\beta}.
\eeqar

\begin{figure}[h]
\begin{center}
\epsfig{file=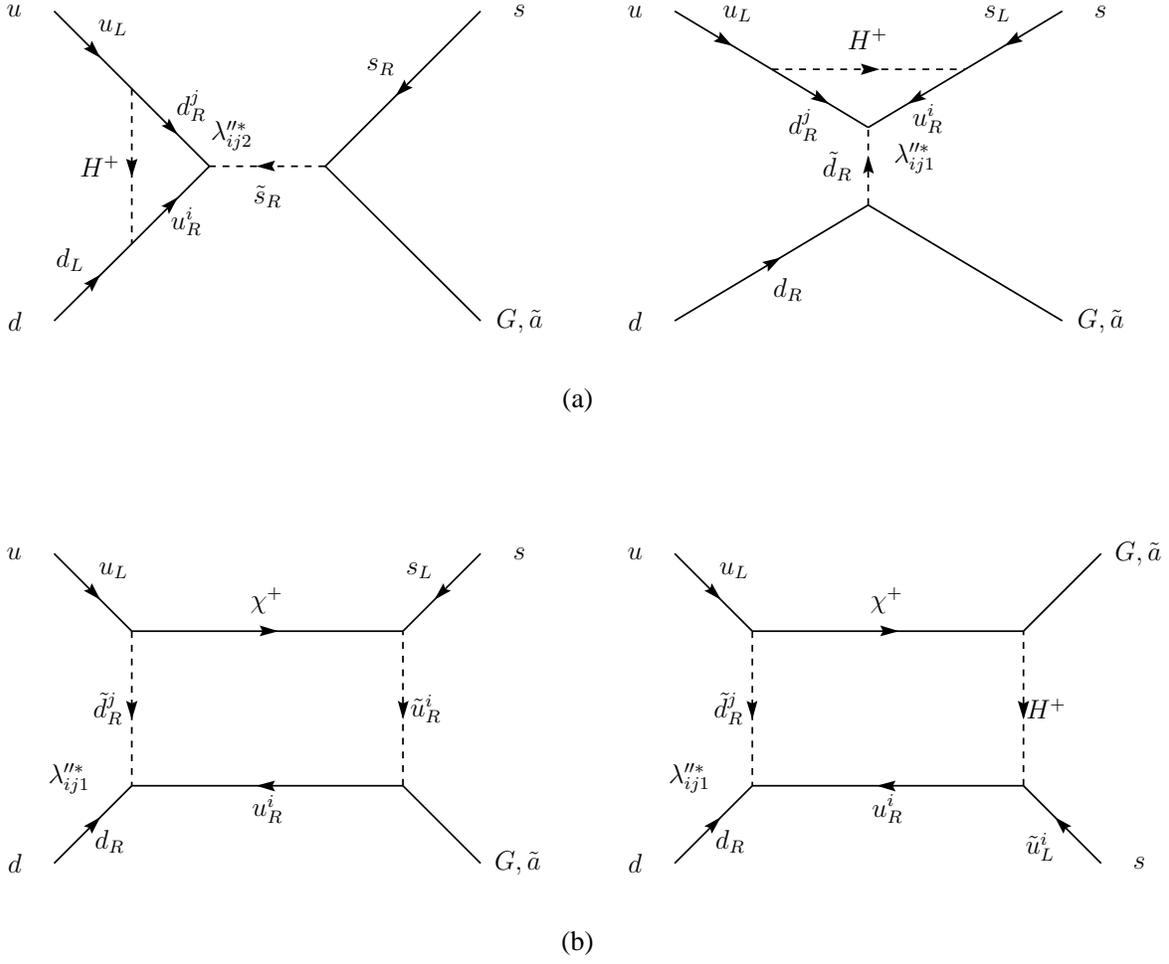,width=16cm}
\end{center}
\caption{
One loop diagrams for  the proton decay into light gravitino
or axino.
}
\label{fig2}
\end{figure}

\begin{figure}[h]
\hspace{1pt}
\begin{center}
\epsfig{file=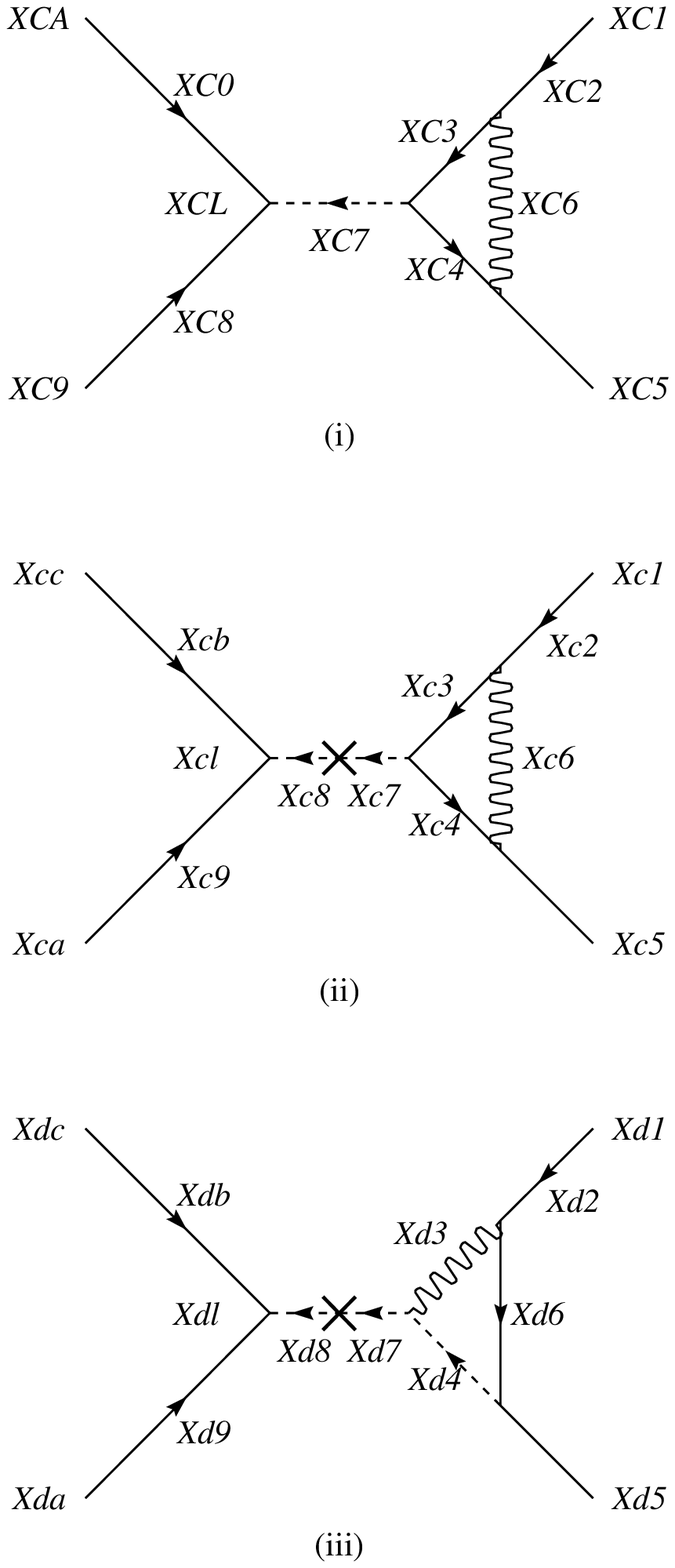}
\end{center}
\caption{
Other one loop diagrams relevant for
$\lambda''_{113}$ and $\lambda''_{123}$.
Here cross means the left-right squark mixing.
}
\label{fig3}
\end{figure}

All one loop diagrams which trigger a proton decay
by having a $\lambda''$-vertex can be
divided into the following three categories; 
(a) diagrams with radiative corrections to the $\lambda''$-vertex 
(Fig.\ref{fig2}),
(b) box diagrams (Fig.\ref{fig2}),  (c) diagrams with radiative
corrections to the gravitino or axino vertex (Fig.\ref{fig3}).
Relative to the tree diagram of Fig.\ref{fig1},
one loop diagrams involving $\lambda''_{ijk}$ will be suppressed
by the factor $\xi_{ijk}$, more explicitly
\beq
\frac{A_{\rm loop}^{(ijk)}}{\lambda''_{ijk}}= 
\xi_{ijk}\frac{A_{\rm tree}}{\lambda''_{112}},
\eeq
where $A_{\rm tree}$  denotes the tree amplitude
of Fig.\ref{fig1}, while $A_{\rm loop}^{(ijk)}$ stand for the loop
amplitudes of 
Fig.\ref{fig2} and Fig.\ref{fig3} which involve the insertion of
$\lambda''_{ijk}$.
The upper bounds on $\lambda''_{ijk}$ resulting from those one loop
diagrams can be easily read off from
(\ref{treeconstraint}) by taking into account the suppression
factor $\xi_{ijk}$:
\beqar
\lambda^{\prime\prime}_{ijk}&\leq& 5\times 10^{-16}
\left(\frac{1}{\xi_{ijk}}\right)
\left(\frac{\tilde{m}}{300 \, {\rm GeV}}\right)^2
\left(\frac{m_{3/2}}{1 \, {\rm eV}}\right) \, , \nonumber \\
\lambda^{\prime\prime}_{ijk}&\leq& 7\times 10^{-16} 
\left(\frac{1}{\xi_{ijk}}\right)
\left(\frac{\tilde{m}}{300 \, {\rm GeV}}\right)^2
\left(\frac{F_a}{10^{10} \, {\rm GeV}}\right) \ 
\left(\frac{1}{c_q}\right) \ . 
\label{loopconstraint}
\eeqar

In the following, we will estimate the size of
$\xi_{ijk}$ for the loop diagrams depicted in Fig.\ref{fig2} and Fig.\ref{fig3}.
Let us first consider 
the type (a) and (b) diagrams in Fig.\ref{fig2}.   
It turns out that 
type (a) diagrams (with the charged Higgs exchange) dominate in this case.
The resulting suppression factors are given by
\beq
\xi_{ijp} 
\approx \frac{1}{(4\pi)^2} \fui V_{iq} \fdj V^*_{1j}=
\frac{g^2}{16\pi^2}\frac{1}{m_W^2\sin(2\beta)} \mui V_{iq} \mdj V^*_{1j},
\label{xia}
\eeq
where $(p,q)=(1,1)$, (1,2) or (2,1).
It is worth noting that these suppression factors are rather insensitive
to the details of unknown superparticle masses.

Although it depends more on the details of superparticle spectrum,
for $\lambda''_{113}$ and $\lambda''_{123}$,
one can get a much stronger bound through the diagrams
in Fig.\ref{fig3}.
For instance, we find that the loop suppresssion factors  of Fig.3--(i)
are given by 
\beqar
\xi_{123} &\approx& 
\frac{g^2}{16\pi^2} V_{31}V_{33}^* \frac{m_b}{m_W}
  \frac{m_t^2 \delta m_{\chi}^2}{\tilde{m}^4} \sin2\phi_L
\approx 5 \times 10^{-7} 
\left(\frac{\delta m_{\chi}^2}{\tilde{m}^2}\right)
\left( \frac{300 \GeV}{\tilde{m}} \right)^2 \sin2\phi_L ,
\nonumber \\
\xi_{113} &\approx& 
\frac{g^2}{16\pi^2} V_{32}V_{33}^* \frac{m_b}{m_W}
  \frac{m_t^2 \delta m_{\chi}^2}{\tilde{m}^4} \sin2\phi_L 
\approx 2\times 10^{-6}
\left(\frac{\delta m_{\chi}^2}{\tilde{m}^2}\right)
\left( \frac{300 \GeV}{\tilde{m}} \right)^2\sin2\phi_L  ,
\label{xic1}
\eeqar
where  we have assumed that all superparticle masses
including the chargino masses are the approximately same as the
universal squark mass $\tilde{m}$,
and  $\delta m_{\chi}^2$ is the
difference between the two chargino mass-squared,
\beq
\delta m_{\chi}^2 \equiv |m_{\chi_1}^2 - m_{\chi_2}^2|. 
\eeq
Here the extra $m_t$-dependence is due to the GIM-cancellation.
In fact, one can consider a diagram which is similar to Fig.3--(i)
but including the insertion of $\lambda''_{212}$
or $\lambda''_{312}$.  However the amplitude of such diagram
is heavily suppressed by the GIM mechanism, and thus it 
does {\it not} give a bound on $\lambda''_{212}$ or $\lambda''_{312}$
which would be stronger than the bound 
from Fig.\ref{fig1}.

If the charginos are  degenerate or the chargino mixing
$|\sin 2\phi_L|\ll 1$, the bounds from
Fig.3--(i)  will be significantly weakened.
In this case, the dominant contribution would come from
Fig.3--(ii) or 3--(iii) which involves the insertion
of the left-right
squark mixing:
\beq
\mdj\mu\tan\beta\tilde{d}^j_L\tilde{d}^j_R+
\mui\mu\cot\beta\tilde{u}^i_L\tilde{u}^i_R+{\rm H.c.}.
\eeq
The corresponding loop  suppression factors  are given by 
\beqar
\xi_{123} &\approx& \frac{g^2}{8\pi^2} V_{31}V_{33}^*
  \frac{\mu m_b}{\tilde{m}^2}
  \frac{m_t^2}{\tilde{m}^2} \approx 4\times 10^{-7} 
\left(\frac{\mu}{\tilde{m}}\right)
\left( \frac{300 \GeV}{\tilde{m}} \right)^3 , \nonumber \\
\xi_{113} &\approx& \frac{g^2}{8\pi^2} V_{32}V_{33}^*
  \frac{\mu m_b}{\tilde{m}^2} 
  \frac{m_t^2}{\tilde{m}^2} \approx 1\times 10^{-6}
\left(\frac{\mu}{\tilde{m}}\right)
\left( \frac{300 \GeV}{\tilde{m}} \right)^3 ,
\label{xic2}
\eeqar
where again it is  assumed that all superparticle masses are
the approximately same as the universal squark mass $\tilde{m}$.

In fact, for the axino case there arises an extra complication for
Fig.3--(i) and Fig.3--(ii) since they involve the axino coupling to 
gauge multiplets ($c_{\alpha}/8\pi^2$ in Eq. (\ref{gacoupling})), 
while the tree diagram of Fig.\ref{fig1} involves only
the axino coupling to the light  quark multiplets ($c_{_I}$  in 
Eq. (\ref{gacoupling})
for $I=q$ where $q=(u,d,s)$ stands for the light quark multiplets).
As a result, for the axino case,
the correct suppression factors of Fig.3--(i) and Fig.3--(ii) are obtained
by multiplying the factor $c_{\alpha}/8\pi^2 c_q$
to the results of Eqs. (\ref{xic1}) and (\ref{xic2}).
This point is irrelevant for the Dine-Fischler-Srednicki-Zhitnitskii type
axino, however it may lead to one order of magnitude stronger bound for
hadronic-type axino. 
In this paper we will ignore this extra complication for the sake of
simplicity.

Applying
the loop suppression factors of Eqs. (\ref{xia}), (\ref{xic1}) and (\ref{xic2})
to  Eq. (\ref{loopconstraint}),
one can easily derive the upper bounds on 
$\lambda''_{ijk}$.
We summarize the numerical results in Table \ref{haha1}.
In deriving these, we take $\sin(2\beta)=1$
in (\ref{xia})   
and ignored the contributions from Fig.3--(i) and Fig.3--(ii) for the
axino case, which would lead to conservative results.
We also assumed that all superparticle masses are
the approximately same as the universal squark mass $\tilde{m}$,
and also  $\mu\approx \tilde{m}$.
For the numerical values of the quark masses, CKM matrix elements and etc.,
the values in Ref. \cite{PDG} are used. 

To conclude, we have examined 
the proton decay involving a light gravitino
or axino in gauge-mediated supersymmetry 
breaking models to derive
constraints on the $R$ parity and baryon number violating
Yukawa couplings $\lambda''_{ijk}$.
Considering the decay amplitudes at one-loop order, we could
get upper bounds on all of those couplings.
The results summarized in Table \ref{haha1} show that, for a wide range of
the gravitino mass $m_{3/2}$, the bounds 
on all $\lambda''_{ijk}$ 
are much stronger than
the currently existing bounds.
The bounds on $\lambda''_{113}$ and $\lambda''_{123}$
are particularly strong due to the contributions from Fig. 3.
In supersymmetric models with $U(1)_{PQ}$, if axino is lighter
than the proton, all $\lambda''_{ijk}$ are similarly constrained
by the proton decay into light axino.

\begin{table}
\caption{\label{haha1}
Upper bounds on $\lambda''_{ijk}$ in gauge-mediated SUSY breaking
models from
the proton decay into  light gravitino (bound I) or axino (bound II). 
Here the bounds on $\lambda''_{113}$ and $\lambda''_{123}$ are from
Fig.\ref{fig3}, while others from Fig.\ref{fig2}.  
All superparticle masses are assumed to be the approximately same
as the universal squark mass $\tilde{m}$, and
$x_s\equiv\left( \frac{\tilde{m}}{300 \GeV} \right)$, $x_{3/2}
\equiv \left( \frac{m_{3/2}}{1 \eV}\right)$, $x_a
\equiv \left( \frac{F_a}{10^{10} \GeV}\right)
\left( \frac{1}{c_q}\right).$
}
%
%

\begin{tabular}{ccc}
Coupling & Upper Bound I & Upper Bound II \\
\hline
$\lambda''_{112}$ & $5\times 10^{-16}$ $x_s^2$ $x_{3/2}$
& $7\times 10^{-16}$ $x_s^2$ $x_a$ \\
$\lambda''_{113}$ & $3\times 10^{-10}$ $x_s^3$ $x_{3/2}$
& $7\times 10^{-10}$ $x_s^3$ $x_a$ \\
$\lambda''_{123}$ & $1\times 10^{-9}$~ $x_s^3$ $x_{3/2}$
& $2\times 10^{-9}$~ $x_s^3$ $x_a$ \\
$\lambda''_{212}$ & $3\times 10^{-8}$~ $x_s^2$ $x_{3/2}$
& $4\times 10^{-8}$~ $x_s^2$ $x_a$ \\
$\lambda''_{213}$ & $5\times 10^{-8}$~ $x_s^2$ $x_{3/2}$
& $7\times 10^{-8}$~ $x_s^2$ $x_a$ \\
$\lambda''_{223}$ & $3\times 10^{-7}$~ $x_s^2$ $x_{3/2}$
& $4\times 10^{-7}$~ $x_s^2$ $x_a$ \\
$\lambda''_{312}$ & $5\times 10^{-9}$~ $x_s^2$ $x_{3/2}$
& $7\times 10^{-9}$~ $x_s^2$ $x_a$ \\
$\lambda''_{313}$ & $1\times 10^{-8}$~ $x_s^2$ $x_{3/2}$
& $1\times 10^{-8}$~ $x_s^2$ $x_a$ \\
$\lambda''_{323}$ & $5\times 10^{-8}$~ $x_s^2$ $x_{3/2}$
& $7\times 10^{-8}$~ $x_s^2$ $x_a$ \\
\hline
\end{tabular}
\end{table}

\psfrag{XAA}[][][0.85]{$u$}
\psfrag{XAB}[][][0.85]{$d$}
\psfrag{XAC}[][][0.85]{$s$}
\psfrag{XA1}[][][0.85]{$u_L$}
\psfrag{XA2}[][][0.85]{$d^j_R$}
\psfrag{XA3}[][][0.85]{$u^i_R$}
\psfrag{XA4}[][][0.85]{$d_L$}
\psfrag{XA5}[][][0.85]{$H^+$}
\psfrag{XA6}[][][0.85]{$\tilde{s}_R$}
\psfrag{XA7}[][][0.85]{$G,\tilde{a}$}
\psfrag{XA8}[][][0.85]{$s_R$}
\psfrag{XAL}[][][0.85]{$\lambda''^*_{ij2}$}

\psfrag{Xaa}[][][0.85]{$u$}
\psfrag{Xab}[][][0.85]{$d$}
\psfrag{Xac}[][][0.85]{$s$}
\psfrag{Xa1}[][][0.85]{$u_L$}
\psfrag{Xa2}[][][0.85]{$d^j_R$}
\psfrag{Xa3}[][][0.85]{$u^i_R$}
\psfrag{Xa4}[][][0.85]{$s_L$}
\psfrag{Xa5}[][][0.85]{$H^+$}
\psfrag{Xa6}[][][0.85]{$\tilde{d}_R$}
\psfrag{Xa8}[][][0.85]{$G,\tilde{a}$}
\psfrag{Xa7}[][][0.85]{$d_R$}
\psfrag{Xal}[][][0.85]{$\lambda''^*_{ij1}$}

\psfrag{XBA}[][][0.85]{$u$}
\psfrag{XBB}[][][0.85]{$d$}
\psfrag{XBC}[][][0.85]{$s$}
\psfrag{XB1}[][][0.85]{$G,\tilde{a}$}
\psfrag{XB2}[][][0.85]{$u_L$}
\psfrag{XB3}[][][0.85]{$\chi^+$}
\psfrag{XB4}[][][0.85]{$s_L$}
\psfrag{XB5}[][][0.85]{$\tilde{u}^i_R$}
\psfrag{XB6}[][][0.85]{$u^i_R$}
\psfrag{XB7}[][][0.85]{$d_R$}
\psfrag{XB8}[][][0.85]{$\tilde{d}^j_R$}
\psfrag{XBL}[][][0.85]{$\lambda''^*_{ij1}$}

\psfrag{Xba}[][][0.85]{$u$}
\psfrag{Xbb}[][][0.85]{$d$}
\psfrag{Xbc}[][][0.85]{$s$}
\psfrag{Xb1}[][][0.85]{$G,\tilde{a}$}
\psfrag{Xb2}[][][0.85]{$u_L$}
\psfrag{Xb3}[][][0.85]{$\chi^+$}
\psfrag{Xb4}[][][0.85]{$H^+$}
\psfrag{Xb5}[][][0.85]{$\tilde{u}^i_L$}
\psfrag{Xb6}[][][0.85]{$u^i_R$}
\psfrag{Xb7}[][][0.85]{$d_R$}
\psfrag{Xb8}[][][0.85]{$\tilde{d}^j_R$}
\psfrag{Xbl}[][][0.85]{$\lambda''^*_{ij1}$}

\psfrag{XC1}[][][0.85]{$d^{\bar{k}}$}
\psfrag{XC2}[][][0.85]{$d^{\bar{k}}_L$}
\psfrag{XC3}[][][0.85]{$t_L$}
\psfrag{XC4}[][][0.85]{$\chi^+$}
\psfrag{XC5}[][][0.85]{$G,\tilde{a}$}
\psfrag{XC6}[][][0.85]{$W^+$}
\psfrag{XC7}[][][0.85]{$\tilde{b}_R$}
\psfrag{XC8}[][][0.85]{$d^k_R$}
\psfrag{XC9}[][][0.85]{$d^k$}
\psfrag{XC0}[][][0.85]{$u_R$}
\psfrag{XCA}[][][0.85]{$u$}
\psfrag{XCL}[][][0.85]{$\lambda''^*_{1k3}$}

\psfrag{Xc1}[][][0.85]{$d^{\bar{k}}$}
\psfrag{Xc2}[][][0.85]{$d^{\bar{k}}_L$}
\psfrag{Xc3}[][][0.85]{$t_L$}
\psfrag{Xc4}[][][0.85]{$\chi^+$}
\psfrag{Xc5}[][][0.85]{$G,\tilde{a}$}
\psfrag{Xc6}[][][0.85]{$W^+$}
\psfrag{Xc7}[][][0.85]{$\tilde{b}_L$}
\psfrag{Xc8}[][][0.85]{$\tilde{b}_R$}
\psfrag{Xc9}[][][0.85]{$d^k_R$}
\psfrag{Xca}[][][0.85]{$d^k$}
\psfrag{Xcb}[][][0.85]{$u_R$}
\psfrag{Xcc}[][][0.85]{$u$}
\psfrag{Xcl}[][][0.85]{$\lambda''^*_{1k3}$}

\psfrag{Xd1}[][][0.85]{$d^{\bar{k}}$}
\psfrag{Xd2}[][][0.85]{$d^{\bar{k}}_L$}
\psfrag{Xd3}[][][0.85]{$W^+$}
\psfrag{Xd4}[][][0.85]{$\tilde{t}_L$}
\psfrag{Xd5}[][][0.85]{$G,\tilde{a}$}
\psfrag{Xd6}[][][0.85]{$t_L$}
\psfrag{Xd7}[][][0.85]{$\tilde{b}_L$}
\psfrag{Xd8}[][][0.85]{$\tilde{b}_R$}
\psfrag{Xd9}[][][0.85]{$d^k_R$}
\psfrag{Xda}[][][0.85]{$d^k$}
\psfrag{Xdb}[][][0.85]{$u_R$}
\psfrag{Xdc}[][][0.85]{$u$}
\psfrag{Xdl}[][][0.85]{$\lambda''^*_{1k3}$}

\newpage
\newpage

\newpage
\hspace{1pt}
\vspace{55pt}

\newpage

\end{document}